\documentclass[useAMS,amssymb,epsfig,usegraphicx,usenatbib,twocolumn,revtex4-1]{emulateapj}

\def\aj{AJ}

\def\apj{ApJ}
\def\apjl{ApJ}
\def\apjs{ApJS}
\def\aap{A\&A}
\def\mnras{MNRAS}
\def\nat{Nature}
\def\na{NewA}
\def\apss{Ap\&SS}
\def\cmda{CeMDA}

\usepackage[usenames,dvips]{color}

\newif\ifAMStwofonts

\makeatother

\shorttitle{Meridional tilt of the stellar velocity ellipsoid}
\shortauthors{Saha et al.}
\begin{document}
\title{Meridional tilt of the stellar velocity ellipsoid during bar buckling instability}

\author {Kanak Saha$^{1}$, Daniel Pfenniger$^{2}$, \& Ronald E. Taam$^{3,4}$}
\affil{$^{1}$ Max-Planck-Institut für Extraterrestrische Physik, Giessenbachstra\ss e, D-85748 Garching, Germany, \\$^{2}$ Geneva Observatory, University of Geneva, CH-1290 Sauverny,
Switzerland,\\ $^{3}$ Institute of Astronomy and Astrophysics, Academia Sinica-TIARA, Taiwan,\\ $^{4}$ Department of Physics and Astronomy, Northwestern University, Evanston, IL 60208, USA \\e-mail: saha@mpe.mpg.de}

\label{firstpage}

\begin{abstract}
The structure and evolution of the stellar velocity ellipsoid plays an
important role in shaping galaxies undergoing bar driven secular evolution
and the eventual formation of a boxy/peanut bulge such as present in the Milky
Way. Using collisionless N-body simulations, we show that during the
formation of such a boxy/peanut bulge, the meridional shear stress of stars,
which can be measured by the meridional tilt of the velocity ellipsoid, reaches
a characteristic peak in its time evolution. It is shown that the onset of
a bar buckling instability is closely connected to the maximum 
meridional tilt of the stellar velocity ellipsoid. Our findings bring new
insight to this complex gravitational instability of the bar which complements
the buckling instability studies based on orbital models. We briefly discuss the
observed diagnostics of the stellar velocity ellipsoid during such a
phenomenon.   
\end{abstract}

\keywords{galaxies: bulges -- galaxies:kinematics and dynamics -- galaxies: structure 
--galaxies:evolution -- Galaxy: disk, galaxies:halos, stellar dynamics}

\section{Introduction}
\label{sec:intro}
Understanding the structure and dynamics of a galaxy crucially depends on the
knowledge of the three dimensional stellar distribution function (DF), which is 
not a direct observable. The first few moments of the DF, e.g., density, mean 
velocity and the velocity dispersion tensor together can provide important clues
regarding the dynamical state of the galaxy and the gravitational instabilities 
it might have undergone \citep{vanderkruit1999}. Of particular interest is the buckling 
instability of a stellar bar in a disk galaxy and the subsequent formation of 
a boxy/peanut bulge \citep{Combesetal1990, PfennigerFriedli1991, Rahaetal1991,
Pfenniger1993,Athanassoula2005}. A bar buckles under its own self-gravity
when it becomes sufficiently strong, thereby bringing substantial changes in the 
velocity distribution of stars and the galactic potential. One possible way to 
quantify such a change is to study the structure and evolution of the stellar 
velocity ellipsoid during the buckling instability and to provide potential diagnostic 
observables. In particular, how is the tilt of the velocity ellipsoid related to
the boxy/peanut bulge such as presented in the Milky Way \citep{Dweketal1995}. This 
requires, however, an unambiguous identification of the buckling event that
a galaxy might be experiencing. However, the onset of buckling instability is not 
clearly understood because it is difficult to follow the orbits of stars subject 
to a rapidly changing gravitational potential during the buckling. 
During this transient phase the dynamics is strongly collective and an 
orbit decomposition can only be a partial description of the process.
Nevertheless, numerous simulation studies marked this event by a decrease in 
the bar strength or in the ratio of vertical-to-radial velocity 
dispersion ($\sigma_z/\sigma_r$) \citep{Combesetal1990, MVetal2006}, providing a 
gross understanding of this event. Simulations show that often such demarcation is 
blurred and a more precise indicator of this event would be useful and complementary 
to the already existing ideas. It is worth re-investigating the buckling instability 
and the relation it might have with the orientation of the stellar velocity ellipsoid, 
in particular with the tilt angle.

The shape and orientation of the stellar velocity ellipsoid are tightly connected
to the symmetry of the underlying galaxy potential \citep{Lindblad1930,
LyndenBell1962, Amendt1991}. In a stationary, axisymmetric disk galaxy the 
stellar velocity ellipsoid in the galactic midplane is perfectly aligned 
with the galactocentric coordinate axes, in other words, all the off-diagonal 
elements of the velocity dispersion tensor are zero \citep{BT1987}. Thus, measuring the 
off-diagonal components of the dispersion tensor in observation may provide one with an 
inference about the presence of non-axisymmetric features in a galaxy. 
Away from the galactic midplane, the tilt of the velocity elliposid might depend on the 
mass distribution of the galactic disc as well as the flattening of the dark matter halo. 
In the context of the Milky Way, the analysis of the RAVE survey data 
release 2 \citep{zwitteretal2008} shows that the velocity ellipsoid is tilted towards 
the Galactic Plane \citep{Siebertetal2008} and has been nicely demonstrated in a recent 
paper by \cite{Pasettoetal2012}. However, the measured tilt angles can not put a strong 
constraint on the disc parameters and halo flattening due to large proper motion errors
 and small sample size in the RAVE DR2 \citep{Siebertetal2008}. 

On the other hand, non-axisymmetric structures such as bars, spiral arms in disk galaxies
might play an important role in accounting for the observed orientation of the stellar
velocity ellipsoid. Numerical study by \cite{vorobyov2008} shows that the vertex deviation 
of the velocity
ellipsoid is globally correlated to the amplitude of the spiral arms.
Using Hipparcos data and dynamical modeling, \cite{Dehnen2000} has shown 
how the Galactic Bar \citep{Blitz1991, Binneyetal1991, Dweketal1995} could have
influenced the velocity distribution in the solar neighborhood. The observed 
low-velocity streams in the solar neighborhood are also thought to have arisen
due to the Galactic Bar \citep{Minchevetal2010}.
It would be useful to understand how the presence of a bar or spiral arms which are highly 
time dependent would change the orientation of the velocity ellipsoid.      

We use N-body simulations to follow the 
evolution of the stellar velocity ellipsoid in a galaxy which undergoes bar 
instability and forms a boxy bulge during the secular evolution in a 
self-consistent way. The buckling of the bar causes the morphological evolution
of the disk, converting its central parts into a boxy/peanut bulge. In order to
gain further insight into the physics of bar buckling \citep{Merrifield1996, MVS2004}, 
we investigate the role of anisotropic stellar pressure and show that there is a 
characteristic signature in the way the stellar velocity ellipsoid evolves. 
The primary goal of this paper is to understand the buckling event of a bar
which forms the boxy/peanut bulge and its relation with the tilt of the stellar 
velocity ellipsoid.

The paper is organized as follows. In the next section, we outline the
general concept of the stellar velocity ellipsoid and the relevant quantities that we
measure from our simulation. Section~\ref{sec:models} briefly describes the
galaxy models used for the present study and simulation. The disk evolution and
boxy bulge formation is described in section~\ref{sec:evolution}. The shear
stress and its relation to bar buckling is shown in
section~\ref{sec:buckling}. We discuss the tilt angle of the velocity ellipsoid
in section~\ref{sec:tilt}. Finally, section~\ref{sec:discus} presents the
discussion and conclusions from this work.          

\section{Stellar velocity ellipsoid}
\label{sec:velep}

The components of the velocity dispersion tensor at a radial location $r$ in the 
stellar disk are computed from the velocity components of a group of stars using 
the following formula \citep{BT1987}:

\begin{equation}
  \sigma^2_{ij} = \langle v_i v_j\rangle - \langle v_i\rangle \langle v_j\rangle,
\end{equation}

where $v_i$ and $v_j$ denote the velocities of a group of stars. $i,j = r,\varphi, z$ 
 in a cylindrical coordinate system. Angular bracket denotes the averaging
over a group of stars. Given the velocity dispersion tensor, the stress tensor of 
the stellar fluid can be written as
\begin{equation}
  \tau = -\rho(r) \sigma^2,
\end{equation}
where $\rho(r)$ is the local volume density of stars at a position $r$. It is 
convenient to think of the entire stress tensor as a sum of two different
kinds of forces acting on a small differential imaginary surface ($dS$) between
two adjacent volumes of stars, i.e., 
\begin{equation}
  \tau = \tau_n + \tau_s, 
\label{eq:stress}
\end{equation}
where $\tau_{n,i} = -\rho(r) \sigma^2_{ii}$ is called the normal stress acting along 
the normal to $dS$ and $\sigma^2_{ii}$ are the diagonal components of the above matrix. 
$\tau_{s,ij} = -\rho(r) \sigma^2_{ij}, i \ne j$ is called the shear stress,
acting along a direction perpendicular to the normal to $dS$, i.e., in the plane
$dS$. In general, the shape of the velocity ellipsoid is determined by the
normal stress, and the shear stress is responsible for the orientation or 
deformation of the ellipsoid w.r.t.\ the galactocentric axes ($\hat{e}_r,
\hat{e}_{\varphi}, \hat{e}_z$). The orientation of the velocity ellipsoid can be
computed using the off-diagonal components of the velocity dispersion tensor.
The meridional tilt of the velocity ellipsoid can be computed using the
following relation:
\begin{equation}
  \Theta_\mathrm{tilt} = \frac{1}{2}\arctan\left[\frac{2 \sigma^2_{rz}}{\sigma^2_{rr} - \sigma^2_{zz}}\right].
\label{eq:etilt}
\end{equation}
We evaluate the shape of the velocity ellipsoid and the tilt angle in the inner region 
of the disk where the dynamics of stars is dominated by a bar and study their evolution as 
the bar enters into the non-linear regime where an analytic understanding is difficult. 

\begin{figure}
\rotatebox{-90}{\includegraphics[height=8.0 cm]{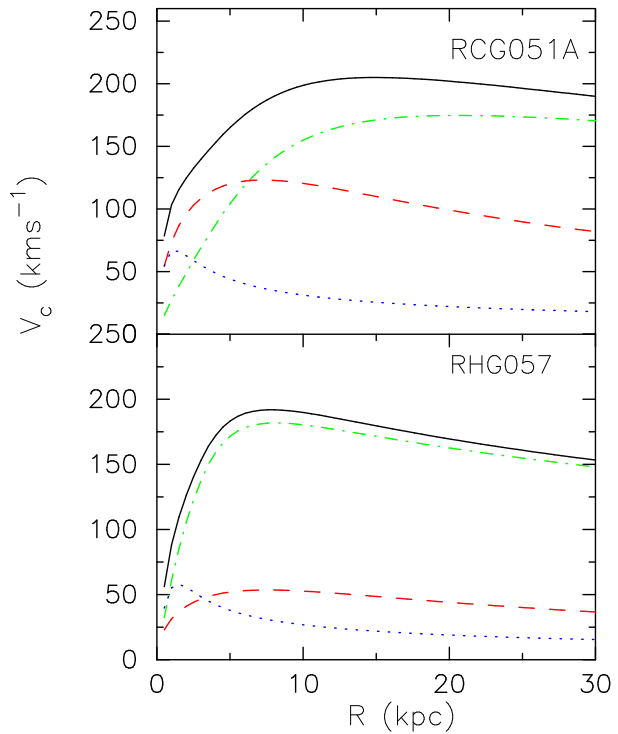}}
\caption{Initial circular velocity curves for the model RCG051A and RHG057. 
In both panels, red dashed line denotes the disk, blue dotted line the bulge and
green dash-dot line the dark halo. Solid black line denotes the total
  circular velocity curve. The inner regions of RCG051A are disk dominated, while
that of RHG057 are dark matter dominated.}
\label{fig:vcall}
\end{figure}

\begin{table}
\caption[ ]{Initial disk, halo and bulge parameters. $Q$ is the Toomre stability 
parameter at $2.5 R_d$; $M_h$ and $M_b$ are the masses of halo and bulge.  }
\begin{flushleft}
\begin{tabular}{lcccccc}  \hline\hline 
  Models    &$Q$ & $M_d$ & $T_\mathrm{orb}$ &$M_h/M_d$ & $M_b/M_d$ \\
            &  & ($\times 10^{10} M_{\odot}$) & (Gyr) & \\
\hline
\hline
RCG051A     & 1.21  & 4.75 & 0.301 & 6.52  & 0.05  \\
RHG097      & 1.84  & 3.11 & 0.306 & 7.88  & 0.43  \\
RHG057      & 2.98  & 0.86 & 0.326& 22.53 & 0.18  \\
\hline
\end{tabular}
\end{flushleft}
\label{tab:paratab}
\end{table}

\section{Initial galaxy models}

\label{sec:models}
In order to study the evolution of the stellar velocity ellipsoid subject to a 
non-axisymmetric bar potential, we perform a large number of simulations of 
isolated galaxies built using the method of \cite{KD1995}. Of these, we present 
here 3 fiducial models (named as RCG051A, RHG057 and RHG097) of disk galaxies
with varying dark matter distribution and Toomre stability parameter (Q).
The initial disk has an exponentially declining surface density with a scale length
$R_d$ and mass $M_d$. The live dark matter halo and bulge are modelled with
a lowered Evans and King DF respectively. For further details on model 
construction, the reader is referred to \cite{Sahaetal2010, Sahaetal2012}. 
We scale the models such that $R_d = 4$~kpc and the disk masses are given in
Table~\ref{tab:paratab}. Orbital time scales $T_\mathrm{orb}$ (at $2.5 R_d$) and other 
initial parameters are given in Table~\ref{tab:paratab}.  
In Fig.~\ref{fig:vcall}, we show the circular velocity curves for
RCG051A and RHG057. Circular velocity curve for RHG097 can be found 
in Fig.~2 of \cite{Sahaetal2010}.    
 
The simulations were performed using the Gadget code \citep{Springeletal2001} 
which uses a variant of the leapfrog method for the time integration. The 
gravitational forces between the particles are calculated using the 
Barnes-Hut tree algorithm with a tolerance parameter $\theta_\mathrm{tol} =0.7$. 
The integration time step used was $\sim 0.82$\,Myr for RCG051A, $0.65$\,Myr for 
RHG097 and $1.5$\,Myr for RHG057. Two of these models were evolved for about 
$6 - 7$~Gyr, and RHG057 was evolved for about 12\,Gyr to understand the long 
term evolution, bar growth and the asymptotic properties of the stellar 
velocity ellipsoid.  

Each of these models were constructed using a total of $2.2$ million particles, 
out of which disk and halo have $1.05$ million each and $0.1$ million 
particles are assigned to the bulge. 
The softening lengths for disk, bulge and halo particles were chosen so that 
the maximum  force on each particle is nearly the same \citep{McMillan2007}. 
In the model RHG097, the softening lengths used for the disk, bulge and halo 
were $12, 25$ and $33$\,pc respectively. For RCG051A, they were $12$, $10$ and 
$31$\,pc and for RHG057, $12$, $17$ and $57$\,pc respectively.
The total energy is conserved well within 0.2\% till the end of the simulation. 

\begin{figure}
\rotatebox{0}{\includegraphics[height=8.5 cm]{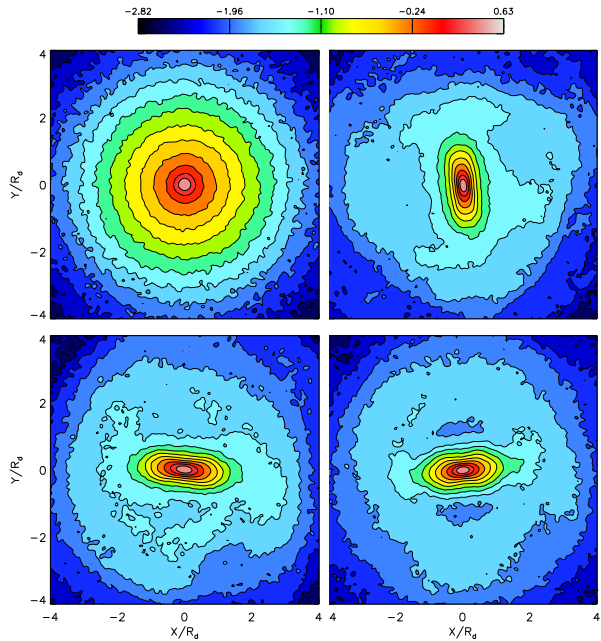}}
\caption{Face-on surface density maps of all the stars belonging to
  the disk and a preexisting classical bulge in the galaxy model
  RCG051A. Top left panel shows surface density at $T=0$, top right at $2.0$, 
bottom left at $3.4$, and bottom right at $5.5$~Gyrs. }
\label{fig:faceon51A}
\end{figure}

\begin{figure}
\rotatebox{0}{\includegraphics[height=8.5 cm]{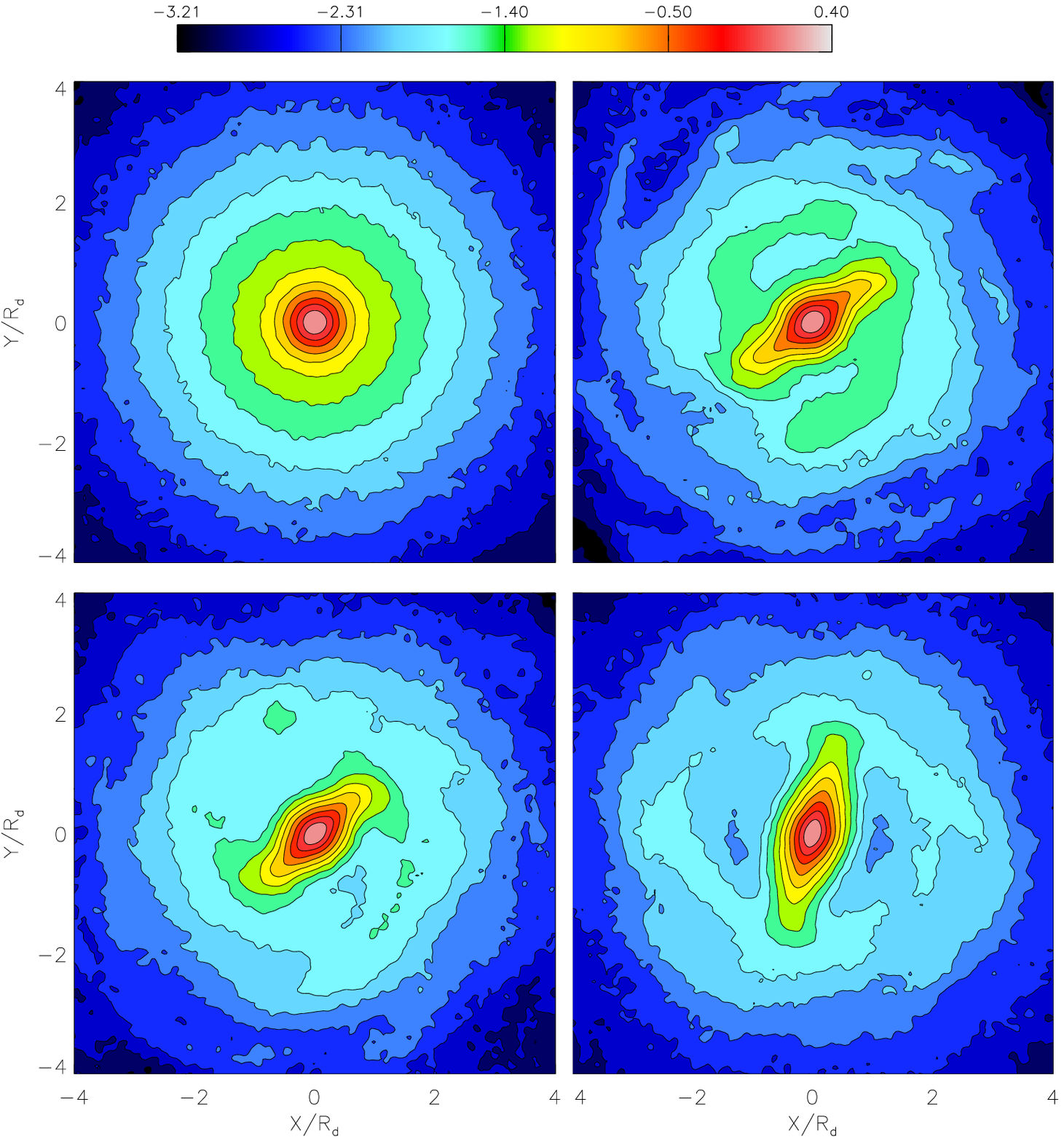}}
\caption{Same as in Fig.~\ref{fig:faceon51A} but for RHG097. Top left panel 
shows surface density at $T=0$, top right at $2.0$, bottom left at $3.4$, 
and bottom right at $5.5$~Gyrs.}
\label{fig:faceon97}
\end{figure}

\begin{figure}
\rotatebox{0}{\includegraphics[height=8.5 cm]{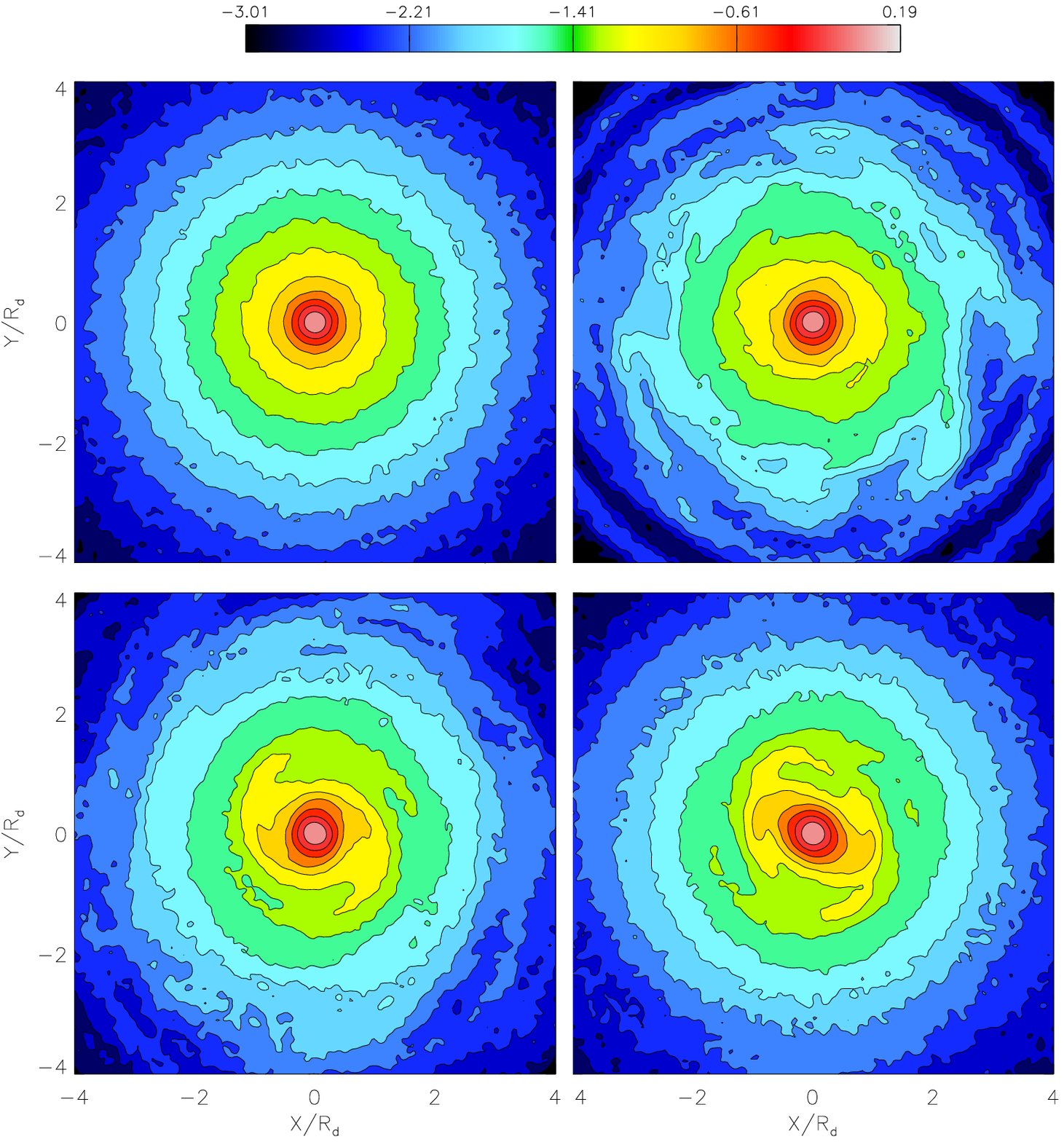}}
\caption{Same as in Fig.~\ref{fig:faceon51A} but for RHG057. Top left panel 
shows surface density at $T=0$, top right at $2.0$, bottom left at $3.5$, 
and bottom right at $7$~Gyrs.}
\label{fig:faceon57}
\end{figure}

\begin{figure}
\rotatebox{-90}{\includegraphics[height=8.5 cm]{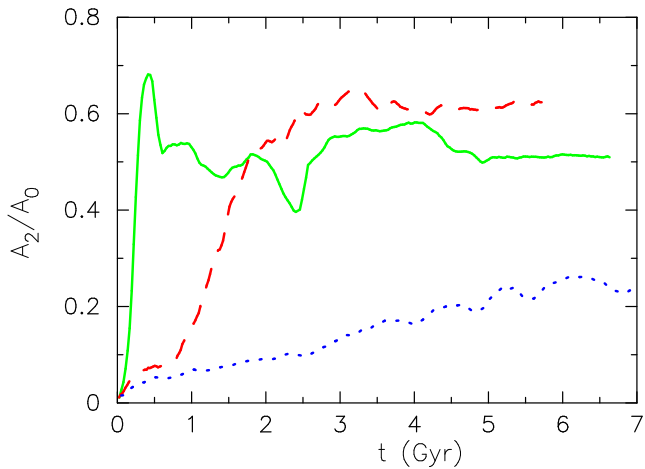}}
\caption{Time evolution of normalized bar amplitude for model RCG051A
  (green), RHG097 (red) and RHG057 (blue).}
\label{fig:A2t}
\end{figure}

\begin{figure}
\rotatebox{-90}{\includegraphics[height=8.0 cm]{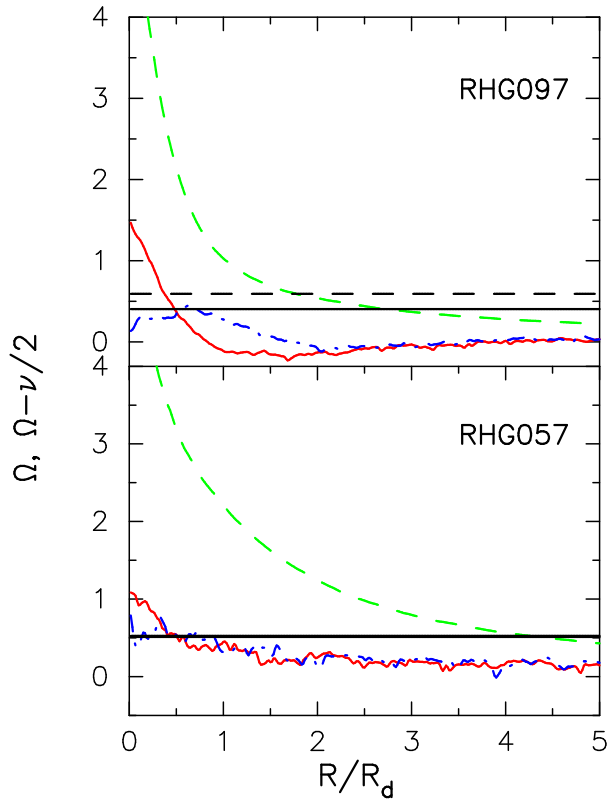}}
\caption{Vertical resonances in the stellar disks of two galaxy models mentioned in 
the figure. Green dashed lines represent the angular frequencies ($\Omega$) 
at $t = 2.2$~Gyr (for upper panel) and $t = 3.6$~Gyr (lower panel). 
In the upper panel, $\Omega -\nu/2$ profiles (red solid and blue dash-dot lines) 
are drawn at $t=2.2$ and $4.8$~Gyr and the corresponding bar pattern speeds at 
those times are denoted by the dashed and solid black lines. 
In the lower panel, they are at $t=3.6$ and $7.0$~Gyr and the corresponding bar pattern 
speeds are denoted by horizontal black lines. The unit of frequencies for RHG097 and 
RHG057 are $44.65$ and $19.0$ kms$^{-1}$kpc$^{-1}$.}
\label{fig:vILR}
\end{figure}

\section{Disk evolution through bar growth}
\label{sec:evolution}

Bar driven secular evolution is an important internal process through which
galaxies change their morphology and kinematics. The rapidity of such a process
depends on various factors of which bar strength plays a significant role. A bar
forms out of the disk instability and grows via nonlinear processes as the disk
stars exchange energy and angular momentum with the surrounding dark matter halo
and a preexisting classical bulge \citep{Sahaetal2012} through gravitational interaction.
The evolution of an initially axisymmetric stellar disk and growth of a bar is
depicted in Fig.~\ref{fig:faceon51A}, Fig.~\ref{fig:faceon97} and
Fig.~\ref{fig:faceon57} which present the surface density maps of all the stars
including that of a preexisting classical bulge.  

In Fig.~\ref{fig:A2t}, we present the time evolution of the bar amplitude measured
by the $m=2$ Fourier component of the surface density of disk stars alone for
the three fiducial models mentioned above. The growth rates of bars are significantly
different in these models which also differ in the relative fraction of dark
matter within the disk region. In model RHG057, the dark matter dominates the
disk right from the center of the galaxy, see Fig.~\ref{fig:vcall}. According to
the classification of \cite{Sahaetal2010}, model RHG057 forms a type-II bar and 
models RCG051A and RHG097 form a type-I bar. Typically, type-I bars are strong and
go through the well known vertical buckling instability \citep{Combesetal1990, 
PfennigerFriedli1991, Rahaetal1991, MVetal2006,Debattistaetal2006} leading to the 
formation of a boxy/peanut (hereafter b/p) bulge as
depicted in Fig.~\ref{fig:edgeon51A} and Fig.~\ref{fig:edgeon97}. Whereas
type-II bars which are weak and grow on secular evolution time scale, normally
do not go through any appreciable buckling instability. We evolved the model
RHG057 for a Hubble time and the disk showed no signature of buckling instability,
although it has grown a moderate size bar by that time. As a result, the
disk in this model did not form a b/p bulge by that time. A thorough understanding of
the buckling instability would perhaps require a tool combining the orbital
analysis and collective effect of the stars in the disk and their role at the
$2:1$ vertical inner Lindblad resonance (ILR). In order to pinpoint the 
location of the ILR, corotation resonances (CR), we first compute the disk frequencies 
($\Omega$, $\kappa$, $\nu$) by a direct sum of the first and second derivatives of
the $N$-body potential obtained from the reflection symmetrized particle distribution with
respect to the $z=0$ plane and the $R=0$ rotation axis. 
We carry out this exercise for each snapshot and compare $\Omega$ and $\Omega-\nu/2$ 
with the pattern speed of the bar. Fig.~\ref{fig:vILR} shows the locations of the 
vertical ILRs and CRs at two different epochs for the two models RHG097 and RHG057. 
In the case of RHG097, the locations of vertical ILRs before ($t = 2.2$~Gyr) and 
after ($t=4.8$~Gyr) the buckling instability are still within the bar region 
indicating that the orbits lie close to the $2:1$ vertical oscillations before and 
after the peanut formation. A detailed orbital analysis by 
\cite{PfennigerFriedli1991} shows that the $2:1$ vertical resonance is essentially
responsible for the formation of b/p bulge, but the collective behaviour of stars
in the vicinity of such resonances remains obscured. For example, it is not
understood what is the role of shear stress or the anisotropic stellar pressure 
in such a process which eventually leads to the formation of b/p bulges. Below we
elaborate on the possible relation between the buckling instability and 
the structure and evolution of the stellar velocity ellipsoid, in particular the 
shear stress due to the disk stars.   

\begin{figure}
\rotatebox{-90}{\includegraphics[height=8.5 cm]{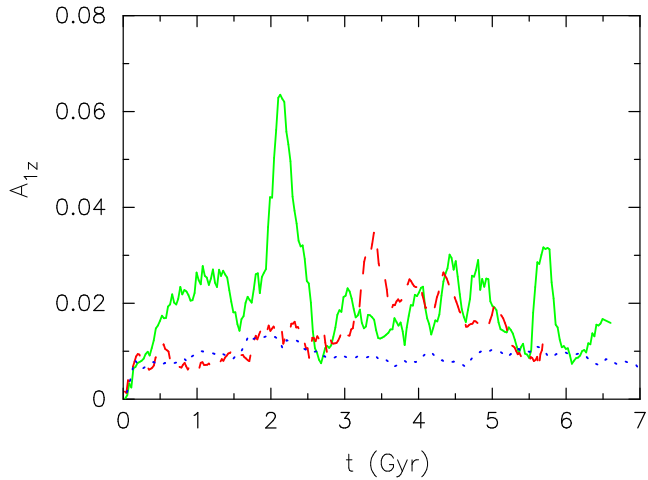}}
\caption{Time evolution of $A_{1,z}$ denoting the vertical asymmetry
  in the bar region. Green line denotes model RCG051A, red line for
  RHG097 and blue for RHG057. }
\label{fig:A1z}
\end{figure}

\section{Shear Stress and buckling instability}
\label{sec:buckling}
As the bar becomes stronger it enters into the regime of the buckling
instability. This instability is highly nonlinear in the sense that the
bending oscillation amplitude is not proportional to self-gravity. In the
case of a bending instability, we would expect a proportionality between the
bending oscillation and the imposed load (here, self-gravity). In the linear
regime, a stellar disk is stable against the low order (e.g., $m=0, 1$) 
bending perturbation and the self-gravity of the perturbation acts like a 
stabilizing agent as shown by several authors, e.g., Toomre (1966), \cite{Araki1985}, 
\cite{Merritt1994}, \cite{Sellwood1996}, \cite{SahaJog2006}. It is the pressure 
forces which destabilizes a stellar disk in response to a bending perturbation. 
From the works of Toomre (1966), \cite{Araki1985} and \cite{Fridman1984}, we 
learnt that a stellar slab of finite thickness would go bending unstable if 
$\sigma_z/\sigma_r < 0.3$. However, the critical value of $\sigma_z/\sigma_r$, at 
which a self-consistent rotating stellar bar would go bending unstable
is unclear. Actually, it is doubtful that a criterion based only on a local 
quantity such as $\sigma_z/\sigma_r$ would apply in a bar, as the $2/1$ vertical 
resonance is a crucial factor which reflects a non-local feature of the system: 
its orbital behavior.  Indeed and contrary to collisional fluids, collisionless 
fluids may develop long range correlations which are not captured by a purely local
description. The distinction between kinematic based and spatial mass distribution
based instabilities in collisionless system has been presented by 
\cite{Pfenniger1996, Pfenniger1998}. A fire-hose instability belongs to 
instabilities depending on a strong gradient in the velocity part
of the DF, while a bar buckling instability belongs to instabilities mainly related
to the presence of a strong resonance, which is determined by the spatial mass
distribution.

In this section, we investigate the role played by the shear stress,
in particular, the meridional component ($\tau_{s,rz}$) which exerts a torque 
in the vertical direction on an imaginary cube of the stellar fluid.        

\begin{figure}
\rotatebox{-90}{\includegraphics[height=8.5 cm]{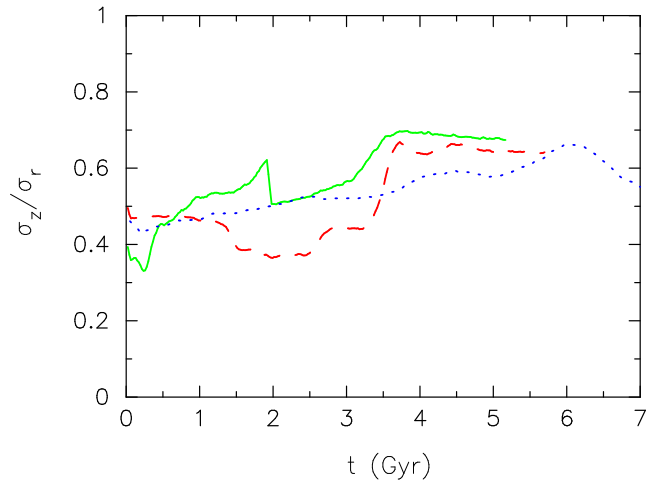}}
\caption{Time evolution of the flattening of the velocity ellipsoid in galaxy
  models RCG051A(green solid line), RHG097 (red dashed line) and RHG057 (blue
dotted line). }
\label{fig:sigzbysigr}
\end{figure}

\begin{figure}
\rotatebox{-90}{\includegraphics[height=8.5 cm]{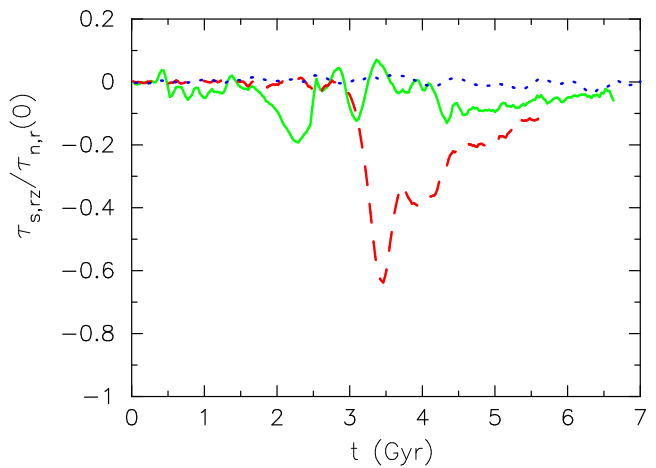}}
\caption{Time evolution of the meridional shear stress normalized by
  the initial normal stress along the radial direction in
  RCG051A(green), RHG097 (red) and RHG057 (blue). }
\label{fig:meridio}
\end{figure}

\begin{figure}
\rotatebox{-90}{\includegraphics[height=8.5 cm]{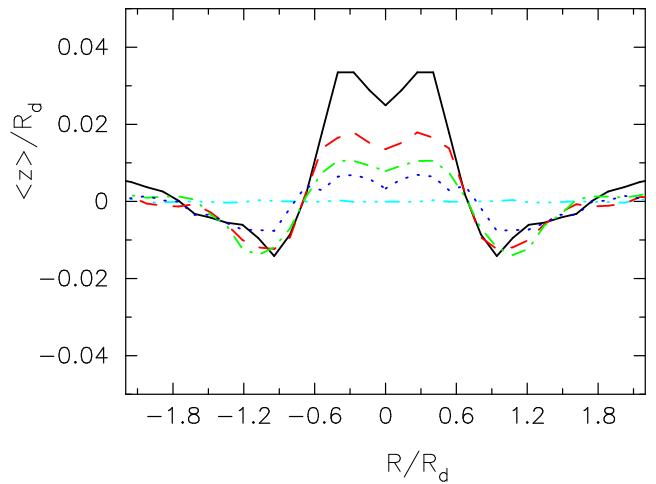}}
\caption{Radial variation of the disk midplane in the bar region for
  the model RHG097. The profiles are drawn at $T=$ 0 (dash-dot-dot line), 3.4 (solid
black line), 4.0 (red dashed line), 4.4 (green dash-dot line) and 5.5 (blue
dotted line) Gyrs.}
\label{fig:meanZ97}
\end{figure}

First, we quantify the buckling amplitude ($A_{1z}$) by computing the $m=1$ Fourier 
component in the $r-z$ plane of the disk with the major axis of the bar aligned to the
disk major axis and identify the buckling phase by studying the time evolution
of $A_{1z}$. During the the buckling phase $A_{1z}$ reaches a peak value and
sometimes goes through a second buckling \citep{MVetal2006}. 
In Fig.~\ref{fig:A1z}, we present the time evolution of $A_{1z}$ for the three
models. $A_{1z}$ for RCG051A shows a strong peak at $T \sim 2$\,Gyr usually
considered as the first buckling of the bar.
It is interesting to note that the bar in this model suffers subsequent buckling of 
smaller amplitudes. The onset of the buckling instability can be indicated by different
physical parameters, e.g., a drop in $A_2$ or $\sigma_z/\sigma_r$ as mentioned
above. For the model RCG051A, both Fig.~\ref{fig:A2t} and Fig.~\ref{fig:sigzbysigr} 
indicate that at around $2$\,Gyr, there is a drop in $A_2$ and $\sigma_z/\sigma_r$
respectively as found in previous studies.

In Fig.~\ref{fig:meridio}, we show the time evolution of the meridional shear stress
normalized by the initial normal stress defined in Eq.~\ref{eq:stress}. It
demonstrates clearly that the meridional stress develop slowly as the bar
evolves and reaches its first peak value as the bar enters the buckling phase at
about $2$\,Gyr following closely the time evolution of $A_{1z}$ in RCG051A.
We establish, here, a new indicator of the bar buckling instability that correlates 
well with other indicators mentioned above in a galaxy with a cold
stellar disk undergoing a rapid phase of bar growth (here, RCG051A). Let us now 
examine the other two models where the bar growth rate is rather slow in 
comparison to RCG051A.

\begin{figure}
\rotatebox{0}{\includegraphics[height=9.3 cm]{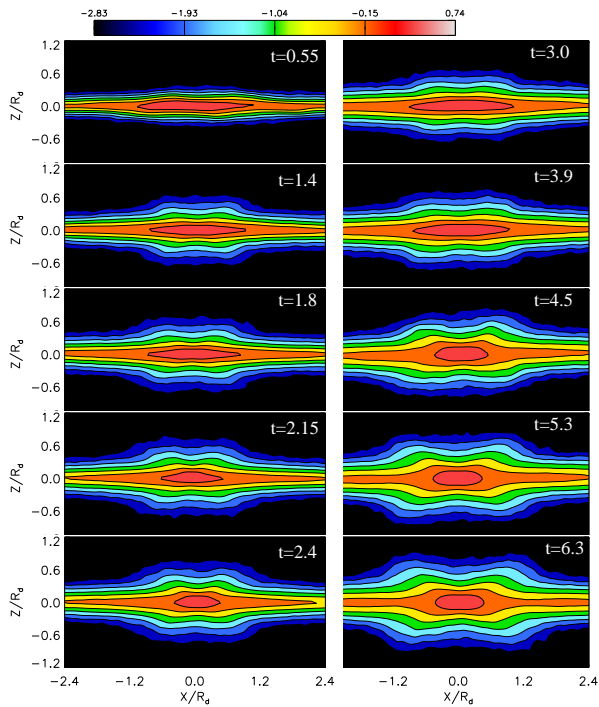}}
\caption{Edge-on projection of the surface density maps of the stellar disk 
in model RCG051A. Time shown in the panels is in Gyr. Buckling instability is 
evident from the vertical asymmetry of the density contours at $t = 2$~Gyrs.}
\label{fig:edgeon51A}
\end{figure}

\begin{figure}
\rotatebox{0}{\includegraphics[height=9.3 cm]{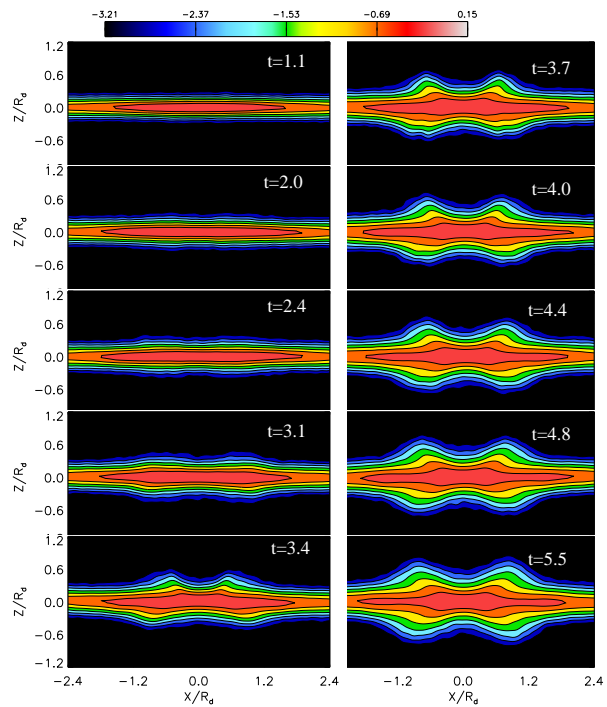}}
\caption{Same as in Fig.~\ref{fig:edgeon51A} but for model RHG097. 
Buckling instability occurs at $t = 3.4$~Gyrs. }
\label{fig:edgeon97}
\end{figure}

\begin{figure}
\rotatebox{0}{\includegraphics[height=9.3 cm]{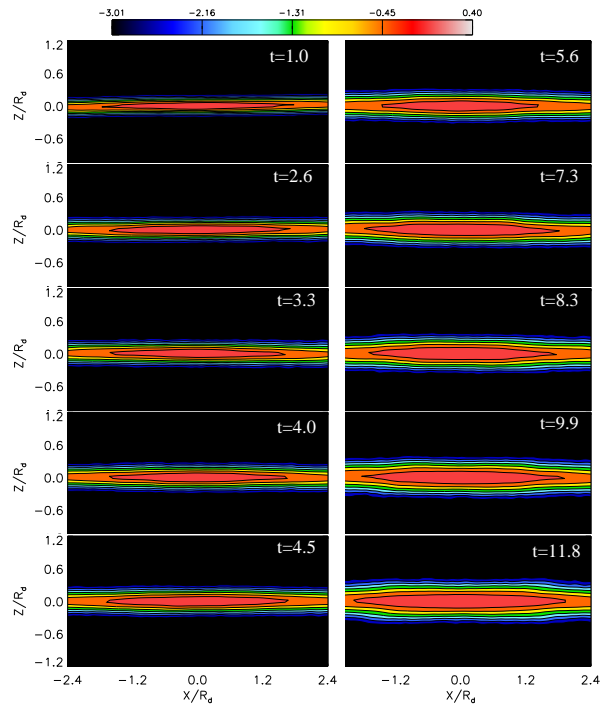}}
\caption{Same as in Fig.~\ref{fig:edgeon51A} but for model RHG057. 
No vertical asymmetry in the density contours detected within $t\simeq12$~Gyrs.}
\label{fig:edgeon57}
\end{figure}

In model RHG097, the peak in $A_{1z}$ (see Fig.~\ref{fig:A1z}) coincides with
that in the meridional stress shown in Fig.~\ref{fig:meridio} at
around $3.4$\,Gyr. However, the drop in $\sigma_z/\sigma_r$ occurs noticeably 
earlier at around $1.5$\,Gyr, when no buckling event was found from the time evolution
of $A_{1z}$ as well as from the visual inspection of the surface density maps 
in edge-on projection (see Fig.~\ref{fig:edgeon97}). {\it It is evident that 
a drop in $\sigma_z/\sigma_r$ is not an unambiguous indicator of the buckling event 
of strong bars, whereas the meridional stress is promising in indicating the 
onset of buckling instability.} 
 
On the other hand, for a dark matter dominated radially hot stellar disk as in 
model RHG057, the bar grows on a much slower rate (see Fig.~\ref{fig:A2t}) and 
shows no peak in $A_{1z}$ (see Fig.~\ref{fig:A1z}). Also Fig.~\ref{fig:sigzbysigr} 
shows no appreciable drop in $\sigma_z/\sigma_r$ except at around $6$\,Gyr where a smooth
decrease in $\sigma_z/\sigma_r$ is apparent. The meridional stress remains nearly 
flat and close to zero for this galaxy model which has been evolved for about 
12\,Gyr during which no buckling event was detected. 

As the bar evolves through the buckling phase, the disk midplane also responds
and exhibits a characteristic buckling mode. We compute the location of the disk
midplane using the following formula to follow the buckling:   

\begin{equation}
\langle z \rangle = \frac{\int {z\rho(r,z) dz}}{\int{\rho(r,z) dz}},
\label{eq:zbar}
\end{equation}

\noindent where $\rho(r,z)$ is the volume density distribution of stars.
Since the meridional shear stress for the model RHG097 was comparatively high,
the disk midplane was expected to show noticeable bending. We discuss here only
the case of RHG097 and mention briefly the other models.

Initially, the disk midplane remains flat at $z=0$ as shown in
Fig.~\ref{fig:meanZ97} for the model RHG097. At $t=3.4$~Gyrs, the midplane
reaches its peak value $\langle z \rangle \sim 130$~pc. Subsequently, the
$z$-amplitude decreases to nearly zero at around $5.5$~Gyrs restoring the symmetry
along the vertical direction. The buckling modes of the bar
in this model have characteristic nodes at $R \sim 0.6 R_d$ and $\sim 1.8 R_d$ 
(see Fig.~\ref{fig:meanZ97}). Comparing with Fig.~\ref{fig:vILR}, we find that the
location of the second node is close to the corotation of the bar.
The time evolution of the $z$-amplitude indicates that the buckling instability 
is a sudden event in the galaxy evolution.  

The model RCG051A also showed similar behaviour in the $z$-amplitude. But the
$z$-amplitude in model RHG057 remained close to zero at all times. 

\begin{figure*}
\rotatebox{0}{\includegraphics[height=24.0 cm]{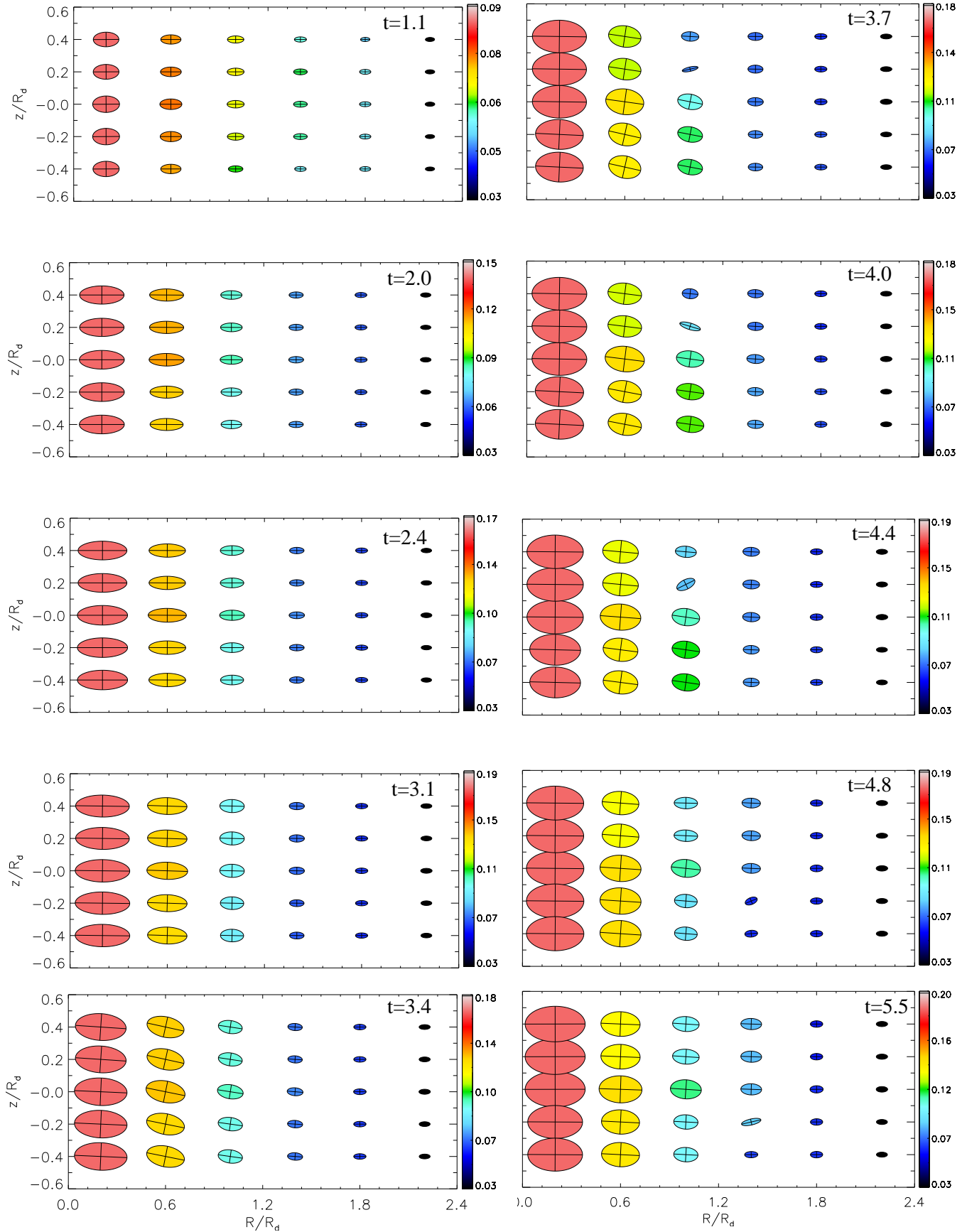}}
\caption{2D map of the stellar velocity ellipsoid in the meridional ($R-z$) plane
 of the disk in model RHG097. Time units are in Gyrs. Color bar represents 
the amplitude of the radial velocity dispersion ($\sigma_r$). The actual value of 
$\sigma_r$ is obtained by multiplying the color indices by $893$ kms$^{-1}$. 
The major and minor axes of the velocity ellipsoid are determined by $\sigma_r$ and 
$\sigma_z$ and they are denoted as inscribed crosses. A rough estimate of the tilt 
angle for each ellipsoid can be gleaned from Fig.~\ref{fig:etilt}. }
\label{fig:2Dtilt}
\end{figure*}

\subsection{Meridional tilt of the velocity ellipsoid}
\label{sec:tilt}
In this section, we discuss the orientation of the stellar velocity ellipsoid in
the meridional plane ($R-z$ plane) of the galactic disk. Since the buckling
instability creates an asymmetry in the vertical density distribution and assuming
it preserves reflection symmetry with respect to the galactic centre, we
consider only one half of the meridional plane for the computation of the
stellar velocity ellipsoid. When the bar has formed, we rotate it so
that its major axis is perpendicular to the line-of-sight; in other words,
the tilt is calculated in the meridional plane containing the bar major axis. In edge-on
projection, the meridional plane would closely resemble the surface density maps
shown, for example, in Fig.~\ref{fig:edgeon97}. In order to understand the
spatio-temporal variation of the stellar velocity ellipsoid in a model galaxy,
we further subdivide the entire meridional plane into several small cells each
of which contain sufficient number of particles for reasonable estimate of the
velocity dispersion and the meridional tilt. The cell sizes are fixed 
at $\Delta R=0.4 R_d$ in radius and $ \Delta z = 0.2 R_d$.  
The number of particles in 
each of these cells vary over time as they are subject to mixing and migration 
driven by the combined effect of an evolving bar and spiral structures in the 
disk \citep{MinchevFamey2010}. To give an idea of the number of particles used, 
the cell at $R = 1.8 - 2.4 R_d$ and $z = 0.3 - 0.5 R_d$ contains about $20,000$ 
particles and the innermost cells have about $100000$ particles at $t = 4.8$~Gyr 
for the model RHG097 (for reference see Fig.~\ref{fig:2Dtilt}).   

Fig.~\ref{fig:2Dtilt} depicts the spatio-temporal variation of the velocity
ellipsoid in the meridional plane of the galaxy model RHG097.
Initially, the velocity ellipsoid in the disk are all aligned with the
galactocentric coordinate axes and the same holds true for a period of about
$3$~Gyr when the bar has fully developed in the disk, see
Fig.~\ref{fig:faceon97}. At $3.4$~Gyr, the bar undergoes a sudden buckling
instability and the meridional tilt of the velocity ellipsoid reaches a peak
value as can be seen from the corresponding panel in Fig.~\ref{fig:2Dtilt}. 
Note that the maximum of the tilt occurs at the first node of the 
buckled bar (see Fig.~\ref{fig:meanZ97}) which roughly coincides with the edge of
the peanut shape in this model (see Fig.~\ref{fig:edgeon97}).
In general, higher values of tilt angle can be found in the b/p region away from 
the minor axis of the galaxy during the buckling phase. It is interesting to 
notice the spontaneous symmetry breaking in the shape distribution of the velocity 
ellipsoid in the meridional plane about the midplane of the galaxy just after the 
peak of the buckling phase. Such asymmetry continues to persist for about 
$1 - 1.5$~Gyr since the onset of buckling instability, during which the density 
distribution is also asymmetric about the midplane (see, Fig.~\ref{fig:edgeon97}). 
After the peak of the buckling phase, the tilt angle of the velocity ellipsoid 
gradually decreases to a low value during the subsequent evolution of the galaxy, 
restoring symmetry both in the shape distribution of the velocity ellipsoid and 
the mass density about the midplane.
    
Note that, the velocity ellipsoid near the minor axis of the galaxy
remains nearly aligned with the galactocentric coordinate axes before the onset
of buckling instability and at later times. Although not shown explicitly here, the
meridional tilt angle of the velocity ellipsoid along the minor axis of the
galaxy remains zero at all times during the galaxy evolution.  
The meridional tilt of the velocity ellipsoid outside the b/p region
is nearly unaffected by the buckling instability. As shown clearly in
Fig.~\ref{fig:2Dtilt} (see panel at $t=2.4$~Gyr), the meridional tilt angle is
nearly zero for galaxies which host a bar that did not go through a buckling
instability. From Fig.~\ref{fig:2Dtilt}, it is clear that the size of 
the velocity ellipsoid near the minor axis nearly doubles at times when buckling 
instability is at its maximum and their sizes continue to increase. Since the semi-major 
axis of the velocity ellisoid actually measures the radial velocity dispersion, it shows 
clear indication of heating in the whole b/p region of the galaxy model.   

\begin{figure}{}
\rotatebox{-90}{\includegraphics[height=8.0 cm]{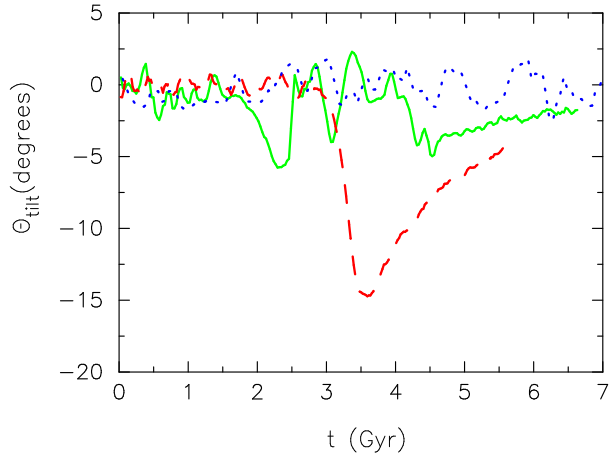}}
\caption{Time evolution of the meridional tilt angle of the stellar
  velocity ellipsoid in the boxy bulge region in three models- RCG051A
  (green), RHG097 (red) and RHG057 (blue).}
\label{fig:etilt}
\end{figure}

Fig.~\ref{fig:etilt} shows the time evolution of the average meridional tilt angle
($\Theta_\mathrm{tilt}$) of the velocity ellipsoid computed in the b/p region for all
the models. In both models RHG097 and RCG051A, the meridional tilt angle rises to a
peak value during the buckling instability phase. On the other hand, the tilt angle
scatters around zero for the model RHG057 at all times. From the time evolution,
a large value of the tilt angle is a characteristic signature of the buckling phase 
that these model galaxies might have undergone. In other words, findings of a large 
value of the tilt angle in the b/p region of a galaxy would indicate that it might be in the 
buckling phase or near the vicinity of the buckling instability. The subsequent evolution 
of the buckling instability in model RHG097 is particularly interesting because of 
the gradual decrement of the tilt angle. It takes about $1$~Gyr for the meridional 
tilt angle to fall by half its peak value and can be considered as the half-life of
the buckling phase ($T_{\mathrm{tilt},1/2}$) the galaxy has experienced. 
We find that $T_{\mathrm{tilt},1/2} \sim 3 \times T_\mathrm{orb}$, where $T_\mathrm{orb}$ 
is the orbital time at the disk half-mass radius (for this model), which is quite short 
compared to the galaxy's lifetime. This might be the reason for the difficulty in 
observing galaxies in the buckling phase. However, $T_\mathrm{tilt,1/2}$ may depend on 
various parameters of the galaxy models and a thorough search of the parameter space is 
required to find an optimal galaxy model which would show large values of tilt angle 
over long periods of time. The dependence of $T_\mathrm{tilt,1/2}$ on the dark halo 
and bulge properties will be considered in a future paper.     

\subsection{Second moment of DF}
\begin{figure}
\rotatebox{0}{\includegraphics[height=14.8 cm]{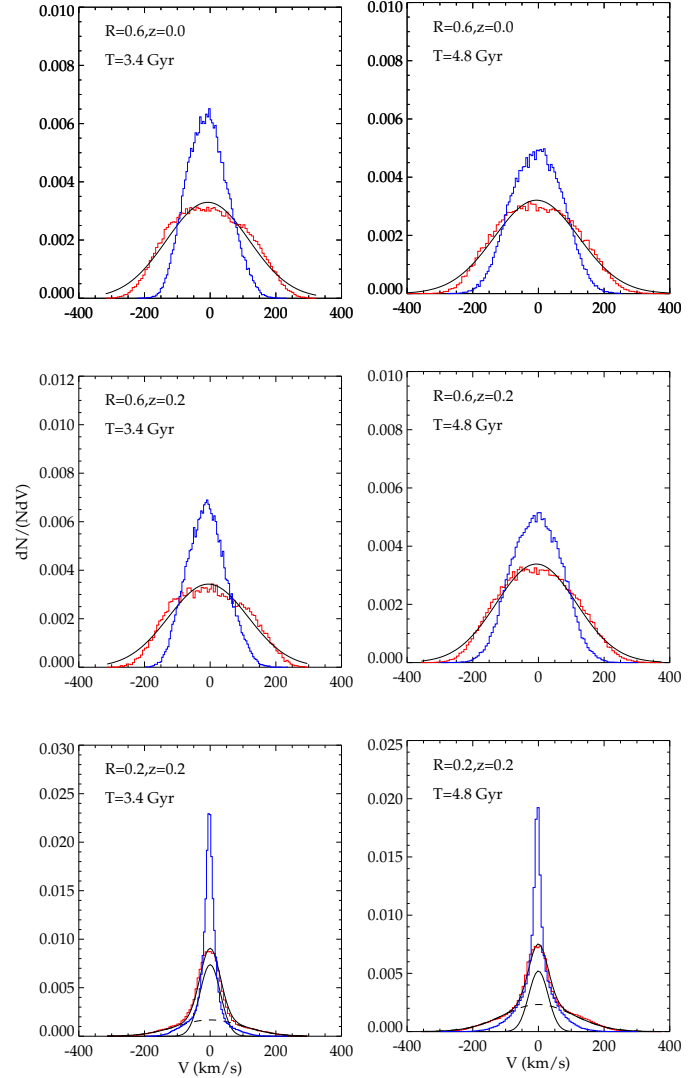}}
\caption{Velocity histograms of stars at $3$ different regions (as indicated 
in the panels) in the meridional plane of the galaxy model RHG097. Blue lines indicate 
vertical velocity and red and broader ones indicate radial. Only the radial velocity 
histograms are fitted with Gaussians, just for illustration.}
\label{fig:df97}
\end{figure}

The calculation of the stellar velocity ellipsoid assumes that the moment integrals
of the DF exist and returns a finite value. However, the validity of such an 
assumption is questionable, especially when the stellar system is undergoing an
unstable phase e.g., buckling instability in the present case. The resonant parts
of the phase space during such an instability can develop bi-modal and/or particle 
distribution with long tail for which the very notion of first or second moment of the DF 
is mathematically no longer meaningful. Bimodal velocity distribution has been observed
for late-type stars in the solar neighborhood by Hipparcos and numerical models of disk 
response to a bar is shown to have reproduced many such features in the local velocity 
distribution \citep{Dehnen2000,Fux2001,Minchevetal2010}.
In Fig.~\ref{fig:df97}, we show the normalized 
histograms for radial ($v_r$) and vertical ($v_z$) velocities in the model RHG097 
during and after the buckling instability nearly disappeared. We picked up three 
different regions in the meridional plane and histograms, in the two regions 
($R=0.6, z=0.$ and $R=0.6, z=0.2$) where the meridional tilt was maximum, are fairly 
well represented by a single Gaussian DF with different variances. In the region close 
to the minor axis of the galaxy i.e., $R=0.2,z=0.2$, the radial velocity histograms needed
two Gaussian DFs: one with cold component with dispersion $\sim 30$ kms$^{-1}$ and one with
a hot component with a dispersion $\sim 102$ kms$^{-1}$. A close inspection of 
Fig.~\ref{fig:2Dtilt} indicates that the stars are heated strongly in that region as it is 
clear from the size of the ellipsoids. At the time of buckling, the size of the ellipsoid 
nearly doubles indicating an increase in the velocity dispersion by a factor of two. In any 
case, in all the regions examined, we have a unimodal DF to represent the stars in the meridional 
plane and they show well behaved first and second moments.   
   
\section{Discussion and conclusions}
\label{sec:discus}
The buckling instability is one of the routes through which an initially axisymmetric
stellar disk would form a boxy/peanut bulge such as present in our own Galaxy.
In order to understand the formation of such a boxy/peanut morphology,
it is important to have further insight on the buckling instability. How and
when would a bar go buckling unstable? How many buckling events has a present day
galaxy experienced? How does it depend on the dark matter
fraction in galaxies? There are several issues needed to be addressed in
order to grasp this phenomenon. The current paper addresses one such issue on the
onset of a buckling event in a disk galaxy. It is shown that there is a connection 
between the onset of the buckling instability and the shear stress of stars. We see that
the shear stress reaches its peak value during the buckling phase and then decreases 
gradually. The development of a shear stress in the stars is a result 
of collective process in the disk. If these stars are also trapped in the
vertical ILRs, this would eventually lead to the buckling instability 
\citep{PfennigerFriedli1991, Quillen2002}.
 
From our study, it is clear that a bar that grows very slowly, on a several Gyr
time scale, does neither develop any appreciable shear stress nor go through any
buckling instability. On the other
hand, bars that grow very rapidly such as in RCG051A, develop both, shear stress and
buckling instability. In no cases that we have studied does a shear stress develop
in the bar and not go through the buckling instability. One emerging scenario is that the
development of shear stress is related to the rate at which a bar grows i.e., 
the rate at which a bar strength grows through transport of angular
momentum outward. In galaxy models with higher values of Toomre $Q$, the growth of bar strength is rather slow leading to weak bar and insignificant amount of shear stress.

Several important kinematical changes occur in the galaxy during and
after the episode of the buckling instability. From the Fig.~\ref{fig:2Dtilt}, it is
clear that the stars are heated in the b/p region, especially close to the minor
axis of the galaxy, by a factor of $\sim 2$ during the
buckling instability. This was shown previously by \cite{Sahaetal2010} for
model RHG097 and others. Another interesting aspect is the kinematical changes in
the vertical structure of the galaxy. Note, the stellar disk is isothermal
initially (see the first panel at $t = 1.1$~Gyr in Fig.~\ref{fig:2Dtilt}).
In the after-episode of the buckling phase, there is a clear distinction in the
velocity dispersion above and below the disk midplane indicating spontaneous
breaking of isothermal structure in the b/p region. Such non-isothermal vertical
structure in the b/p region is persistent long after the buckling phase. We will
address this issue in more detail in a future paper.  

Our main conclusions from this work are as follows:

\noindent 1. We show that the meridional tilt of the stellar velocity ellipsoid is a
{\it better indicator} compared to a drop in the bar amplitude or $\sigma_z/\sigma_r$ for the onset of the buckling instability of a stellar bar in a
disk galaxy. During the buckling event, the tilt angle reaches a peak value
followed by a gradual decrease. Outside the buckling episode, the tilt angle is
nearly zero. The meridional shear stress of stars and the onset of the
buckling instability of a stellar bar is closely connected. 

\noindent 2. A large value of the tilt angle of the stellar velocity ellipsoid 
in the b/p region indicates
the occurrence of a buckling event in the galaxy. The meridional tilt angle of
the velocity ellipsoid remains close to zero if the bar does not experience  the
buckling phase.   
 
\noindent 3. Disk galaxies that are radially hot and highly dominated by the dark matter
halo might not have gone through a buckling instability. Buckling instability
appears to depend on the growth rate of bar strength.              

\noindent 4. Buckling instability changes the vertical structure and kinematics in the
boxy/peanut region of the galaxy, in particular it changes vertical structure from isothermal to non-isothermal in one of our models.
\vspace{-0.25cm}
\section*{Acknowledgement}
\noindent  The simulations presented in this paper were carried out on the computer cluster 
system of ASIAA. This work was supported, in part, by the Theoretical Institute for Advanced Research in Astrophysics operated under the ASIAA. K.S. acknowledges support from the
Alexander von Humboldt Foundation. D.P. acknowledges support from the Swiss
National Science Foundation. The authors thank the anonymous referee for insightful comments on the manuscript.


\end{document}